\documentclass[aps,prx,twocolumn,superscriptaddress,showpacs,floatfix]{revtex4-2}
\usepackage{color}
\usepackage{mathtools}
\usepackage[c]{esvect}
\usepackage{bm}       
\usepackage{amssymb}   
\usepackage{amssymb, amsthm, float, graphicx, amsfonts, dsfont}
\usepackage{amsmath} 

\usepackage[colorlinks=true,citecolor=blue,linkcolor=magenta, breaklinks=true]{hyperref}
\usepackage{graphicx}
\usepackage{braket}
\usepackage{natbib}
\usepackage{blkarray, bigstrut}
\usepackage[dvipsnames]{xcolor}
\usepackage{comment}
\usepackage{physics}
\usepackage{tikz}
\usepackage[english]{babel}
\usepackage[utf8]{inputenc}

\begin{document}
% \title{Probing octupolar order through impurity-induced strain}
%TC:ignore
%\begin{document}

\title{Exploring magnetic and topological complexity in MgMn$_6$Sn$_6$: from frustrated ground states to nontrivial Hall conductivity}

%Flat band separation featuring a robust spin Hall Effect with nontrivial band topology in the presence of finite correlation in bilayer kagome metals.} 

\author{Jyotirmoy Sau}
\affiliation{Department of Condensed Matter and Materials Physics,
S. N. Bose National Centre for Basic Sciences, JD Block, Sector III, Salt Lake, Kolkata 700106, India}

\author{Hrishit Banerjee}
\email{hbanerjee001@dundee.ac.uk}
\affiliation{School of Science and Engineering, University of Dundee, Scotland, UK}
\affiliation{Yusuf Hamied Department of Chemistry, University of Cambridge, Cambridge, UK}

\author{Sourabh Saha}
\affiliation{Department of Condensed Matter and Materials Physics,
S. N. Bose National Centre for Basic Sciences, JD Block, Sector III, Salt Lake, Kolkata 700106, India}

\author{Nitesh Kumar }
\email{nitesh.kumar@bose.res.in}
\affiliation{Department of Condensed Matter and Materials Physics,
S. N. Bose National Centre for Basic Sciences, JD Block, Sector III, Salt Lake, Kolkata 700106, India}
\author{Manoranjan Kumar}
\email{manoranjan.kumar@bose.res.in}
\affiliation{Department of Condensed Matter and Materials Physics,
S. N. Bose National Centre for Basic Sciences, JD Block, Sector III, Salt Lake, Kolkata 700106, India}

%\date{\today}
\begin{abstract}
 We explore the intriguing topological itinerant magnet MgMn$_6$Sn$_6$, characterized by bilayer kagome Mn layers encasing a hexagonal Sn layer. Using \textit{ab initio} Density functional theory and Dynamical mean-field theory calculations, we uncover the complex electronic properties and many-body configuration of its magnetic ground state. Mn d-orbital electrons form a frustrated many-body ground state with significant quantum fluctuations, resulting in competing antiferromagnetic and ferromagnetic spin exchanges. Our band dispersion calculations reveal a mirror symmetry-protected nodal line in the \textit{k}$_z$ = 0 plane. When spin-orbit coupling (SOC) is introduced, the gap is formed along the nodal line lifted due to broken time-reversal symmetry with magnetic ordering, leading to substantial intrinsic Berry curvature. We identify Dirac fermions, van Hove singularities, and flat band near the Fermi energy (\textit{E}$_F$), with SOC introducing a finite gap at key points. The unique proximity of the flat band to  \textit{E}$_F$ suggests potential instabilities. Spin-orbit coupling opens a 20 meV gap at the quadratic touching point between the Dirac and flat band, bestowing a nonzero Z$_2$ invariant. This leads to a significant spin Hall conductivity. Despite the presence of large incoherent scattering due to electronic interactions, band crossings and flat band features persist at finite temperatures. MgMn$_6$Sn$_6$ exhibits intriguing topological and magnetic properties, with promising applications in spintronics.
\end{abstract}
%\pacs{75.25.aj, 75.40.Gb, 75.70.Tj}
\maketitle
%TC:endignore
%\section{Introduction}
The kagome lattice,  formed of vertices and edges of the trihexagonal tiling pattern,  exhibits geometrical frustration due to its corner-sharing triangles and transition element based magnets on this geometry is intriguing  due to their tendency to display correlated topological band structures \cite{yin1,nakatsuji,ye,liu,kang,lin}. The topologically protected quantum states of the strongly correlated systems, particularly on non-trivial geometries like the kagome lattice, hold significant promise for exploring exotic quantum phases\cite{qp,qp1}, and their diverse application in quantum technologies\cite{keimer,sachdev,hasan,he,franz,armitage,yin,thouless,haldane,chang,sharpe,zou,tang,xu}. The electronic band structure of the kagome lattice shows the destructive interference of Bloch wavefunctions which results in non-dispersive flat bands (FB)\cite{kang,kang_1,mazin,xu}, Dirac cone at high symmetry point \textit{K} \cite{dirac,dirac1} and van Hove singularity (VHS) at \textit{M} point\cite{vhs}. The instabilities near the flat band states have the potential to induce emergent novel phenomena, including ferromagnetism, high-temperature superconductivity, and the fractional quantum Hall effect\cite{tasaki,mielke,peotta,imada,wu,tang,neupert}. 

Recently, the Hubbard Hamiltonian\cite{model,model1} has been employed on a Kagome lattice to understand the electronic properties and emergence of novel phases predicted theoretically \cite{novel,novel2} and observed experimentally \cite{novel1} in the system. The spinless Haldane model is another commonly employed model to understand the electronic properties of structure in kagome lattices which incorporates SOC and out-of-plane ferromagnetism \cite{haldane}. It predicts the emergence of Chern gap at Dirac points (DP)\cite{ye,yin}, a phenomenon observed in various kagome magnets. Recent experimental findings have corroborated the theoretical predictions, revealing spin-polarized Dirac cones with SOC induced gaps in two-dimensional (2D) kagome materials such as Fe$_3$Sn$_2$\cite{ye,wang,yin_1} and Co$_3$Sn$_2$S$_2$\cite{liu,liu_1,wang_1,yin_2}, as well as the manifestation of anomalous Hall effects in compounds such as Mn$_3$Sn and Mn$_3$Ge\cite{nayak,Naoki} due to gigantic Berry curvature(BC).

It is possible to envisage layer stacking of 2D kagome materials forming bilayer three-dimensional (3D) kagome materials. The \textit{A}Mn$_6$Sn$_6$(\textit{A} = Li,Mg,Ca) family is such a set of bilayer 3D Kagome materials that have recently attracted considerable interest due to the exhibition of many interesting electronic properties like the presence of Chern-gapped Dirac fermions in TbMn$_6$Sn$_6$, which also demonstrates a ferrimagnetic order perpendicular to the kagome lattice plane. YMn$_6$Sn$_6$ \cite{asaba,chen_1,wang_2,ma} exhibits an intriguing topological Hall effect near room temperature. However, emergent properties origination from the FB are often not observed (Fe$_3$Sn$_2$, Co$_3$Sn$_2$S$_2$)\cite{wang,lin,hasan} or far away from \textit{E}$_F$ (FeSn)\cite{kang}. The VHS and flat bands found in previously studied Mn-, Fe-, and Co-based kagome metals are typically situated far from the E$_F$ in energy space. However,  the material with FB near \textit{E}$_F$ is most desirable due to it's easy tunability.  

 Unlike the model kagome lattice, the dispersion of the FB in a bilayer system can be modified by various types of factors such as in-plane next-nearest-neighbour hopping, interlayer coupling, and multiple orbital degrees of freedom of transition metals \cite{next,next1}. In the case of the rare-earth based \textit{A}Mn$_6$Sn$_6$ or related systems, atoms of the rare earth sit close to Mn-kagome layer and they influence the electronic band structure significantly \cite{play}. Therefore, the experimental realization of the kagome flat band requires careful and systematic material design. These complexities along with the interplay between electron correlation and topological features, pose challenges in detecting the FBs, saddle points, and Dirac fermions near the \textit{E}$_F$ in magnetic kagome systems. 

In this work, we examine the bilayer Kagome material MgMn$_6$Sn$_6$  consisting of two kagome Mn single layers sandwiching the MgSn layer
where non-magnetic Mg atom sits in the Sn plane and does not interact with the electrons Mn kagome layer. We notice from our minimal multiband Hubbard model that Sn atoms play a crucial role in determining the nature of magnetic exchange in the Mn layer. We use ab initio density functional theory (DFT) calculations to explore the electronic properties of MgMn$_6$Sn$_6$. The energy band structure of this system shows many interesting properties: First, the Dirac point is located at the Brillouin zone (BZ) corner and just below the \textit{E}$_F$. Second,  a nodal line exists at the \textit{k}$_z$ = 0 plane, which opens a narrow gap in the presence of SOC which also generates finite BC along this line. It exhibits an intrinsic anomalous Hall effect (AHE) due to the non-trivial band topology, with an anomalous Hall conductivity $\sigma^{A}_{xy}$ reaching a significant value of 500 S/cm. Another intriguing observation is the opening of a gap at the touching point of the quadratic band, emerging from the Dirac band, and flat bands in the presence of SOC. The nontrivial topology of the gapped-out flat band results in a finite spin Hall conductivity (SHC) and this is confirmed by calculating the $Z_2$ index. We also discussed how the Mn d-orbital electrons form a frustrated many-body ground state with significant quantum fluctuations, resulting from competing antiferromagnetic and ferromagnetic spin exchanges. We also perform dynamical mean field theory (DMFT) calculations to include the effects of electronic correlations in the system and show that the estimated Curie temperature $T_c \sim 300$ K is consistent with magnetocaloric experiment\cite{expt,expt1}.
\begin{figure*}[t]
    \centering
    \includegraphics[width=\textwidth]{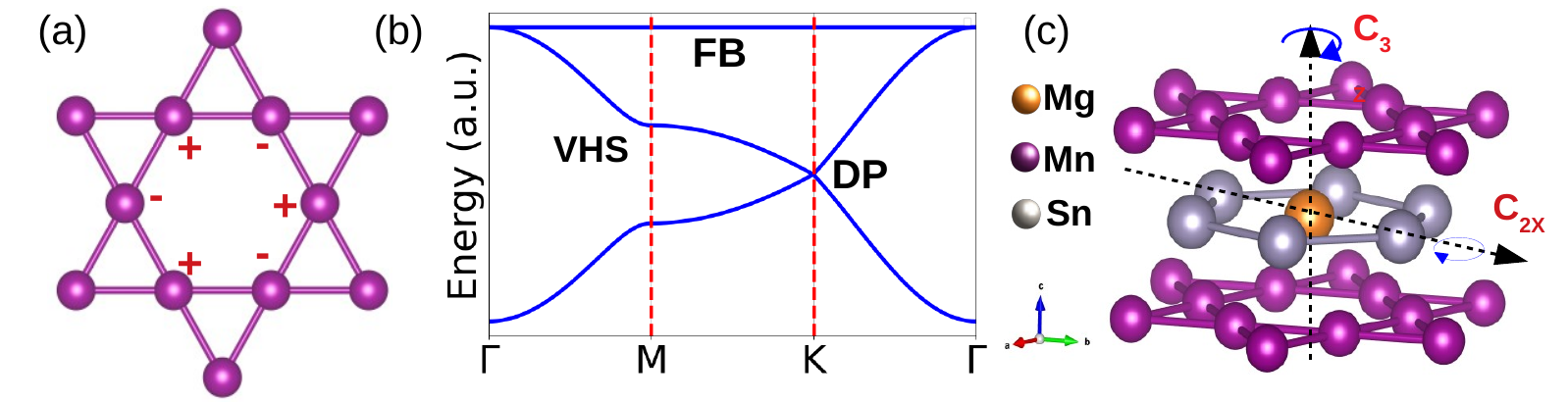}
    \caption{(a) Structure of
kagome lattice and a quenched eigenstate induced by destructive interference. (b) The
band structure of electronic kagome lattice without SOC, with NN interaction dominating the
in-plane hopping process.  (c) The kagome metal's three-fold (\textit{C$_{3z}$}) and two-fold(\textit{C$_{2x}$}) rotation symmetries are when the two kagome layers sandwich the hexagonal Sn layer.}
    \label{fig1}
\end{figure*}

\begin{figure*}[t]
    \centering
    \includegraphics[width=\textwidth]{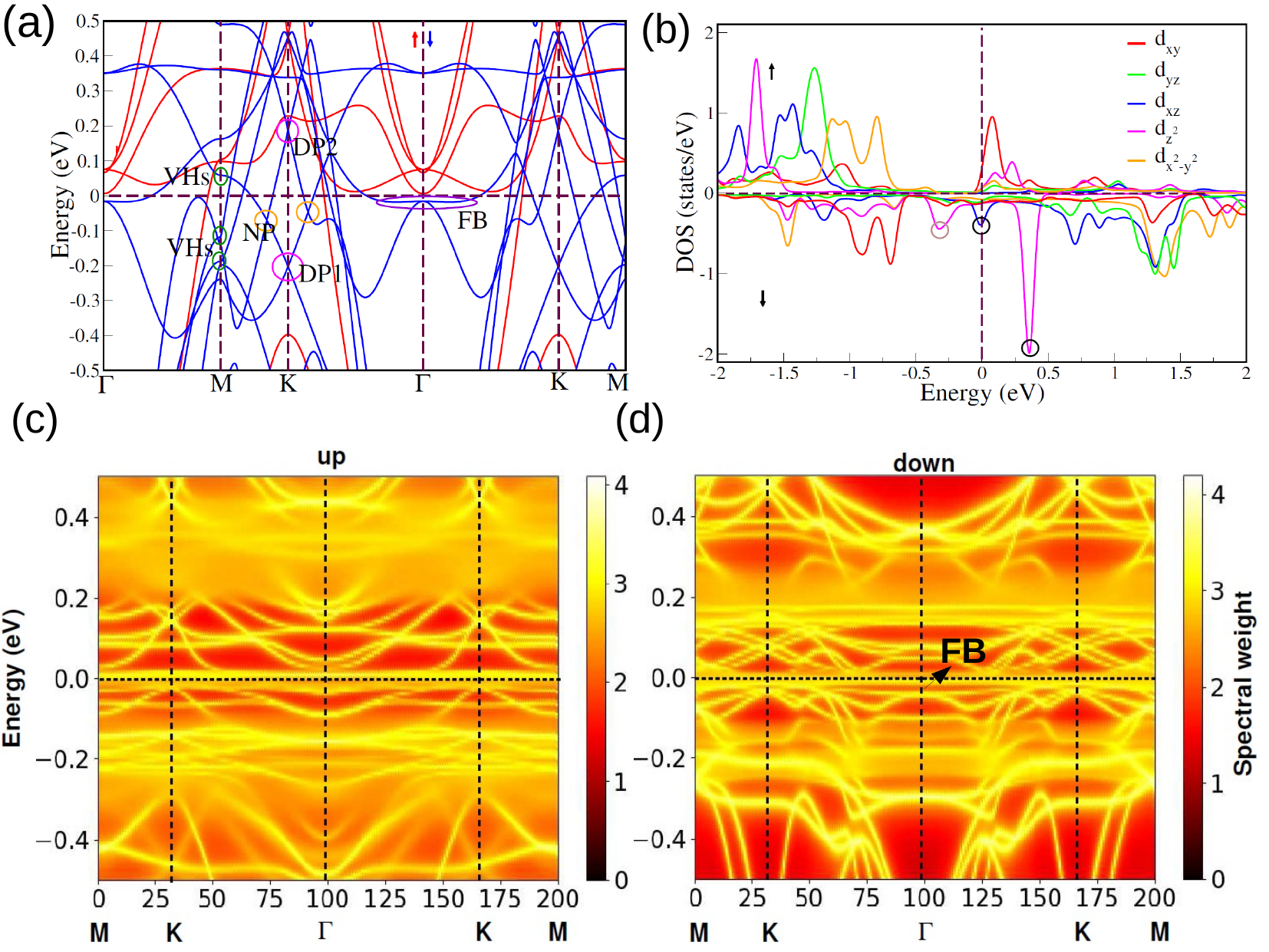}
    \caption{(a) The band structures, excluding spin-orbit coupling (SOC), depict majority and minority spin carriers with red and blue curves, respectively. Notable features such as the FB, DP1 and DP2,  VHS, and Weyl points are denoted by maroon, magenta, green, and orange circles, respectively. (b) The DOS of different d-orbitals of the Mn atom displays a metallic character. The peak in the minority DOS
is attributed to FB1, FB2, and SP. (c),(d) DMFT bands at $\beta$=150eV$^{-1}$ for the \textit{K} path \textit{M}-\textit{K}-$\Gamma$-\textit{K}-\textit{M}. This figure shows the effect of incoherent scattering due to dynamic correlation effects at finite temperature on the band structure in the spin-polarised state while maintaining the same features of the bands as observed with 0K DFT. }
    \label{fig2}
\end{figure*}

\section{\label{sec2}Crystal Structure}
The kagome compound MgMn$_6$Sn$_6$ possesses a HfFe$_6$Ge$_6$-type hexagonal structure, falling within the space group \textit{P6/mmm}, with lattice parameters of a = b = 5.517 \text{\AA} and c = 9.032 \text{\AA} and $\alpha$ = $\beta$ = 90 $^{\circ}$ and $\gamma$ =120 $^{\circ}$ . It comprises double Mn Kagome lattice layers parallel to the ab plane, with a honeycomb lattice of Sn atoms.

\section{Results}

\subsection{ab initio calculations}

The electronic band structure of a single-orbital kagome lattice (see Fig.\ref{fig1}(a)) in the tight binding paradigm is initially explored, as illustrated in Fig.\ref{fig1}(b).
%This electronic state is confined geometrically within a single hexagon because any hopping to neighboring cells is impeded by destructive phase interference, as depicted in Fig.\ref{fig1}(a). This localization in real space translates into momentum-space (Bloch) eigenfunctions exhibiting no energy dispersion, resulting in flat bands, as shown in Fig.\ref{fig1}(b). 
In this limit, the flat band emerges alongside a pair of Dirac cones and VHS which arise due to the fact that the honeycomb lattice and the Dirac points are protected by lattice symmetry \cite{dirac,dirac1,vhs}. The flat band arises due to the destructive interference of electron wavefunction and the phase of the wavefunctions as shown in Fig.\ref{fig1}(a) \cite{flat}. 

MgMn$_6$Sn$_6$ is a more complex system where the honeycomb Sn layer is sandwiched between two kagome layers as depicted in Fig.\ref{fig1}(c) and this requires a first principles approach for the electronic dispersion calculation.
Let us then analyze the electronic band dispersion of our system where the spin-up and -down bands are shown in blue and red colour (see Fig.\ref{fig2}(a)) as obtained from PBE+U calculations. These bands exhibit linear crossing points near the \textit{E}$_F$ without SOC and it is shown in $\textit{k}_z$=0 plane along $\Gamma$-\textit{M}-\textit{K}-$\Gamma$-\textit{K}-\textit{M}, illustrated in Fig.\ref{fig2}(a). 

The first striking feature in the band structure is the presence of Dirac cones with linear dispersion that are observed at the corner points \textit{K} and \textit{K}$^\prime$ of the BZ, owing to the protection provided by two-fold and three-fold rotational symmetries\cite{c3z} (C$_{2x}$ and C$_{3z}$, respectively) of the kagome layer, as depicted in Fig.\ref{fig1}(c). The magenta circle in Fig.\ref{fig2}(a) highlights the location of the band's Dirac point1 (DP1) at the \textit{K} point, situated approximately 0.25 eV below \textit{E}$_F$. Orbital-projected band calculations indicate that the Dirac point primarily originates from the out-of-plane Mn d$_{z^2}$ orbital (magenta), as illustrated in Fig.S1(b). Another Dirac point 2 (DP2), indicative of the band structure emerging from the kagome lattice, is identified at the \textit{K} point, placed approximately 0.19 eV above \textit{E}$_F$ and created by the Mn atoms d$_{z^2}$ and d$_{x^2-y^2}$ orbitals. Furthermore, we have illustrated the constant energy contours at DP1 and DP2, revealing the distinctive hexagonal symmetry of the kagome lattice. Additionally, a circular electron pocket is identified near the center of the Brillouin zone ($\Gamma$), as depicted in Fig.S2. 

Next, we observe in Fig.\ref{fig2}(a) a degenerate crossing point positioned below the \textit{E}$_F$, marked by an orange circle, which is identified as a Weyl point. Further detailed analysis of this observation is provided in part C. Additionally, a linear band crossing point is noted along the high-symmetry \textit{M}-\textit{K} direction at the \textit{E}$_F$, denoted by a black circle in Fig.\ref{fig2}(a), which forms the nodal-line-like band structure in momentum space. part C will provide a specific analysis of these nodal lines in more detail. 
%The estimated Dirac velocities are as follows: v$_{DP1}$ = 11.25 $\times$ 10$^5$ ms$^{-1}$ and v$_{DP2}$ = 19 $\times$ 10$^5$ ms$^{-1}$. 

We also observe the presence of VHS in the energy dispersion of the tight-binding model on the Kagome lattice, as illustrated in Fig.\ref{fig1}(b). MgMn$_6$Sn$_6$ however exhibits two types of VHS with opposite concavities: the m-type VHS which demonstrates an upward energy shift as the band approaches the \textit{M} point and the p-type VHS which exhibits a downward energy shift as the band approaches the \textit{M} point\cite{van,van1}. These findings have been shown in Fig.S1(d). Additionally, this VHS phenomenon signifies a divergence in the density of states (DOS), as illustrated in Fig.\ref{fig2}(b).

Finally, the most important observation in the band structure of MgMn$_6$Sn$_6$, is the flat band(FB) which has d$_{xz}$ and d$_{yz}$ orbital character of Mn atoms. This FB feature appears along the $\Gamma-K$ direction at 0.04 eV below E$_F$. The dispersion of the FB in this material may be influenced by many factors beyond spin-orbit coupling, such as in-plane next-nearest-neighbor hopping, interlayer coupling, or multiple orbital degrees of freedom as discussed in Sec.V of the supplemental section. 

Next, we move on to analyse the partial density of states (PDOS) of Mn atoms near the \textit{E}$_F$ and notice that the major contribution to the DOS comes from d-orbitals of the Mn as shown in Fig.\ref{fig2}(b). The peak in the DOS in the minority spin channel marked by the black circle ( Fig.\ref{fig2}(b)), confirms that the Mn minority spin channels are responsible for the formation of FB. At the $\Gamma$ point, a quadratic band (QB) emerges from the Dirac band (DP2) and touches the FB, as depicted in Fig.\ref{fig2}(a). This touching point remains robust against perturbations except for SOC. Further details on the tight-binding calculation can be found in Sec.V in the supplemental section. Both the flat band and the quadratic band are primarily composed of d$_{xz}$/d$_{yz}$ orbitals of Mn, as illustrated in Fig.S1(b). 

%To date, several ARPES studies have reported the observation of kagome lattice-derived flat bands in materials like GdV$_6$Sn$_6$, YMn$_6$Sn$_6$, CoSn, Fe$_3$Sn$_2$, and FeSn\cite{}. However, there is limited clear evidence regarding the kagome-derived flat band in the 135 families represented by AV$_3$Sb$_5$ (A = K, Rb, or Cs). 

Further investigation of the robustness of this touching point requires exploration of the impact of dynamic correlations and finite temperature effects on the band dispersion which are completely missed by standard DFT+U calculations. We have conducted this analysis by DFT+DMFT calculations. In contrast to DFT+U which represents a static correlation energy correction without frequency dependence, DFT+DMFT includes both static and dynamic correlation effects within the frequency dependent self energies. Utilizing continuous-time Quantum Monte Carlo (QMC) solvers within hybridization expansion, we simulate the influence of temperature. We observe the effect of dynamic correlations through the finite frequency self-energy and utilize this self-energy to compute momentum- and energy-resolved correlated spectral functions, as illustrated in Fig.\ref{fig2}(c). These spectral functions represent a DMFT band structure incorporating both correlation effects from the self-energy and finite temperature effects, providing a direct comparison with ARPES. This scattering predominantly exhibits incoherent scattering characteristics, evident from the large values of Im $\Sigma$ for $\omega \rightarrow 0$ (frequency ($\omega$)-dependent dynamic self-energy), as shown in Fig.S5. This behavior aligns with previous findings in near half-filled manganites \cite{lmo-cat} and other strongly correlated materials \cite{bro}. While dynamic correlations and finite temperature effects introduce some mass renormalization, the features observed at 0K remain preserved.

\begin{figure*}[t]
    \centering
    \includegraphics[width=\textwidth]{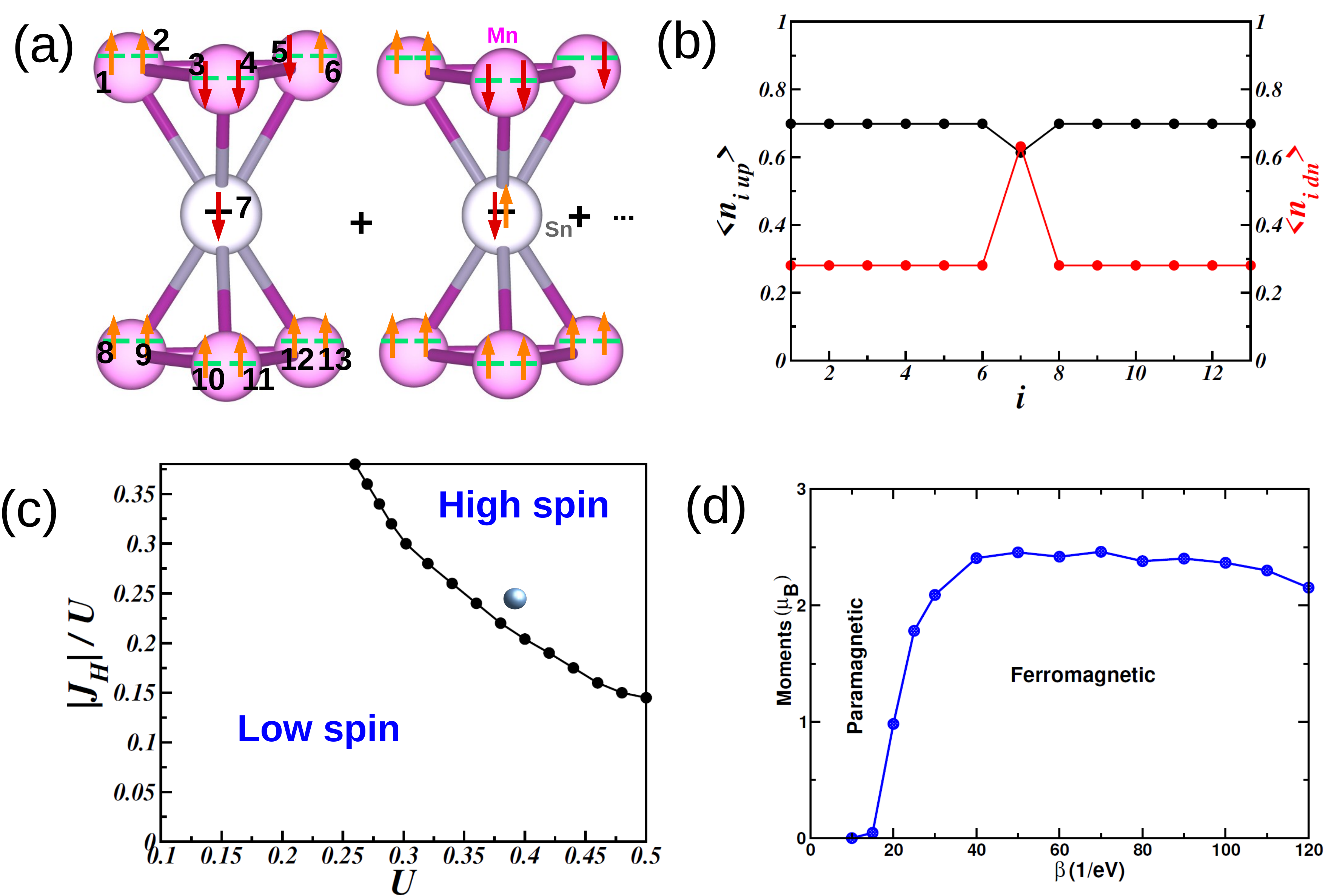}
    \caption{(a) The unit cell of a bi-layer kagome structure featuring one Sn atom, along with the predominant configurations in the ground state wave function. (b) The average charge density for an up spin and down spin within the unit cell, where the black curve represents the up spin charge density and the red curve represents the down spin charge density. (c) The phase diagram for the unit cell as a function U and J$_H/U$, with the blue solid ball representing the appropriate parameter for MgMn$_6$Sn$_6$. (d) The evolution of magnetic states as a function of temperature.}
    \label{fig3}
\end{figure*}

\subsection{Model}

We now turn our attention to the magnetic properties of the material, aiming to understand the correlated nature of the ground state using a many-body model Hamiltonian. The density of states (DOS) and atomic-projected band structure reveal that the electronic dispersion near \textit{E}$_F$  is primarily due to the d-orbitals of Mn and the p-orbitals of Sn, as shown in Fig.S1. In the total DOS, the spin-up DOS has a lower value at \textit{E}$_F$ compared to the spin-down DOS, while the spin-up DOS is higher at lower energies. This indicates the localization of the spin-up d-orbitals of Mn. The local moment of Mn is measured by DFT+U to be 2.3 $\mu_{B}$/Mn, consistent with experimental measurements \cite{expt}. Bader charge analysis\cite{b1,b2} suggests that the d-orbitals of Mn and the p-orbitals of Sn are half-filled.

All the d-orbitals of Mn are equivalent, and similarly, the three p-orbitals of Sn are also equivalent. Therefore, we construct a minimal multi-band Hubbard model, including only two d-orbitals and two electrons of each Mn and one p-orbital and one electron of each  Sn, to study the emergence of the magnetic ground state within the relevant parameter space. We consider a unit cell of the bi-layer kagome system, as depicted in Fig.\ref{fig3}(a). This includes six Mn atoms and one Sn atom. We use a multiband Hubbard model which includes electronic interaction or many body terms H$^{int}_{i}$ and kinetic or electron hopping term $H^{tb}_{ij}$. So, the total model Hamiltonian can be written as 
\begin{eqnarray}
H=\sum_{i}H^{int}_{i}+\sum_{<ij>}H^{tb}_{ij},
\end{eqnarray}
where the interaction term on i$^{th}$ multi-orbital site can be written written  as:
\begin{eqnarray}
H_i^{int} =  \sum_{i, \alpha} U_i 
 n_{i\uparrow \alpha}n_{i\downarrow \alpha} +  \sum_{i, \alpha < \beta} (U_i' - \frac{J_{i,H}}{2}) n_{i \alpha}n_{i \beta} \nonumber \\
 - 2  \sum_{i, \alpha < \beta} J_{i,H}  \bold{S}_{i \alpha} . \bold{S}_{i \beta}+\sum_{i,\alpha}\Delta_{i,\alpha}n_{i,\alpha} \hspace{1cm}. 
\end{eqnarray}

The first two terms are the intra and inter-orbital Hubbard interaction at $i$ $^{th}$site. The third term is the inter-orbital Hund's coupling between the spins at site $i$. $U_i$, $U_i'$ and $J_{i,H}$ are the usual Kanamori parameters for i$^{th}$ site and we use the standard relation $U_{i}' = U_{i} - 2J_{i,H}$.  For Mn, we assumed the Coulomb interaction, $U$, and the Hund's coupling, $J_{H}$ as 0.4 and 0.25 respectively\cite{parameter}but in the case of Sn, these parameters are taken to be 0. $\Delta_{i,\alpha}$ represents the chemical potential for the $\alpha$ orbital of i$^{th}$ site. For the Sn atom, it is -1.5 eV and for the e$_{g}$ orbitals of Mn, it is taken as 0 eV, as suggested by Wannier calculation.
Now, the kinetic energy of electron between i$^{th}$ and j $^{th}$ site $H^{tb}_{ij}$ can be  written as
\begin{eqnarray}
H^{tb}_{ij} = \sum_{ \sigma, \alpha, \beta} t_{i,j,\alpha, \beta} (c^{\dagger}_{i\sigma \alpha}c_{j\sigma \beta} + H.c.), 
\end{eqnarray}
where, $t_{i,j,\alpha, \beta}$ is the hopping strength of electron between the orbital $\alpha$ of $i$-th site and orbital $\beta$ of $j$-th site. $c^{\dagger}_{i\sigma \alpha}$ is the creation operator of an electron with spin $\sigma$, orbital $\alpha$, at site $i$ whereas, $c_{j\sigma \beta}$ is the annihilation operator with the same spin at site $j$, orbital $\beta$. The hopping interaction is restricted to the two $d$ orbitals of different Mn atoms and one $p$ orbital of Sn. The hopping matrix element of $t_{\alpha-\beta}^{Mn-Mn}$ and $t_{\alpha}^{Mn-Sn}$ are calculated based on the wannier calculations and given in in Sec.III in the supplemental section. We solve this Hamiltonian, given in eq.(1) in many body basis using exact diagonalization (ED) technique.
The hopping of electrons from Mn to Mn are either direct or through the Sn atom, leading to effective antiferromagnetic and ferromagnetic exchange respectively. These two competing interactions lead to frustration in the system. The effective spin of the frustrated ground state of the structure is 5/2 and it is twelve-fold degenerate (6$\times$2), where 6 arises from spin degeneracy (2S+1=6)  and each of these states are doubly degenerate due to inversion symmetry about the axis passing through the Sn atoms.  We notice that all orbitals of Mn atoms are singly occupied due to the onsite Coulomb repulsion (U).  In Fig.\ref{fig3}(a), we illustrate the two most dominant configurations in the ground state wave function.  The density of spin up, $<n^{\uparrow}_{i}>$ and spin down, $<n^{\downarrow}_{i}>$ are shown in Fig.\ref{fig3}(b) for orbitals/sites and numbering of the orbitals $i$ is shown in Fig.\ref{fig3}(a). The average spin density 
$<S_i^z>=1/2(<n^{\uparrow}_{i}>$-$<n^{\downarrow}_{i}>$) for each orbital of Mn is 0.21 and for Sn p orbital, it is -0.02.  Therefore, the average spin density per Mn is 1.05 assuming the g=2, and this is consistent with the experimental reported data \cite{expt}.  This type of ferromagnetic ground state is also observed in the rare earth bi-layer compounds such as \textit{R}Mn$_6$Sn$_6$ (where \textit{R} = Li, Mg, Ca, Tb, Ho, Er, Tm, Lu, and Dy). The magnetic ground state (gs) with  S=5/2 of the model Hamiltonian is a function of Hubbard $U$ and $J_H$,  therefore, the magnetic gs phase diagram is constructed as a function of both the parameters of the Mn atom for a given value of hopping terms. The magnetic phase diagram is shown in Fig.\ref{fig3}(c). It is observed that the gs of this system goes from a low spin (S=1/2) state to a high spin (S=5/2) state by tuning U and J$_H$ as shown in Fig.\ref{fig3}(c). The solid sphere represents the relevant parameter of MgMn$_6$Sn$_6$. The low spin state S=1/2 has an antiferromagnetic alignment of spins between the different orbitals of the Mn site. Further details of the S=1/2 magnetic gs are discussed in the supplemental section.  

We also examine the evolution of the magnetic states with temperature using DMFT calculations, as illustrated in  Fig.\ref{fig3}(d). Initially, at high temperatures, a paramagnetic state is observed. As the temperature is reduced, ferromagnetic ordering begins to emerge around $\beta=20eV^{-1}$ (corresponding to \textit{T}=580K), reaching a saturation magnetization of approximately $2.45 \mu_B$ at $\beta=40eV^{-1}$ (\textit{T}=290K), which aligns excellently with the experimental findings\cite{expt,expt1}.

\begin{figure*}[t]
    \centering
    \includegraphics[width=\textwidth]{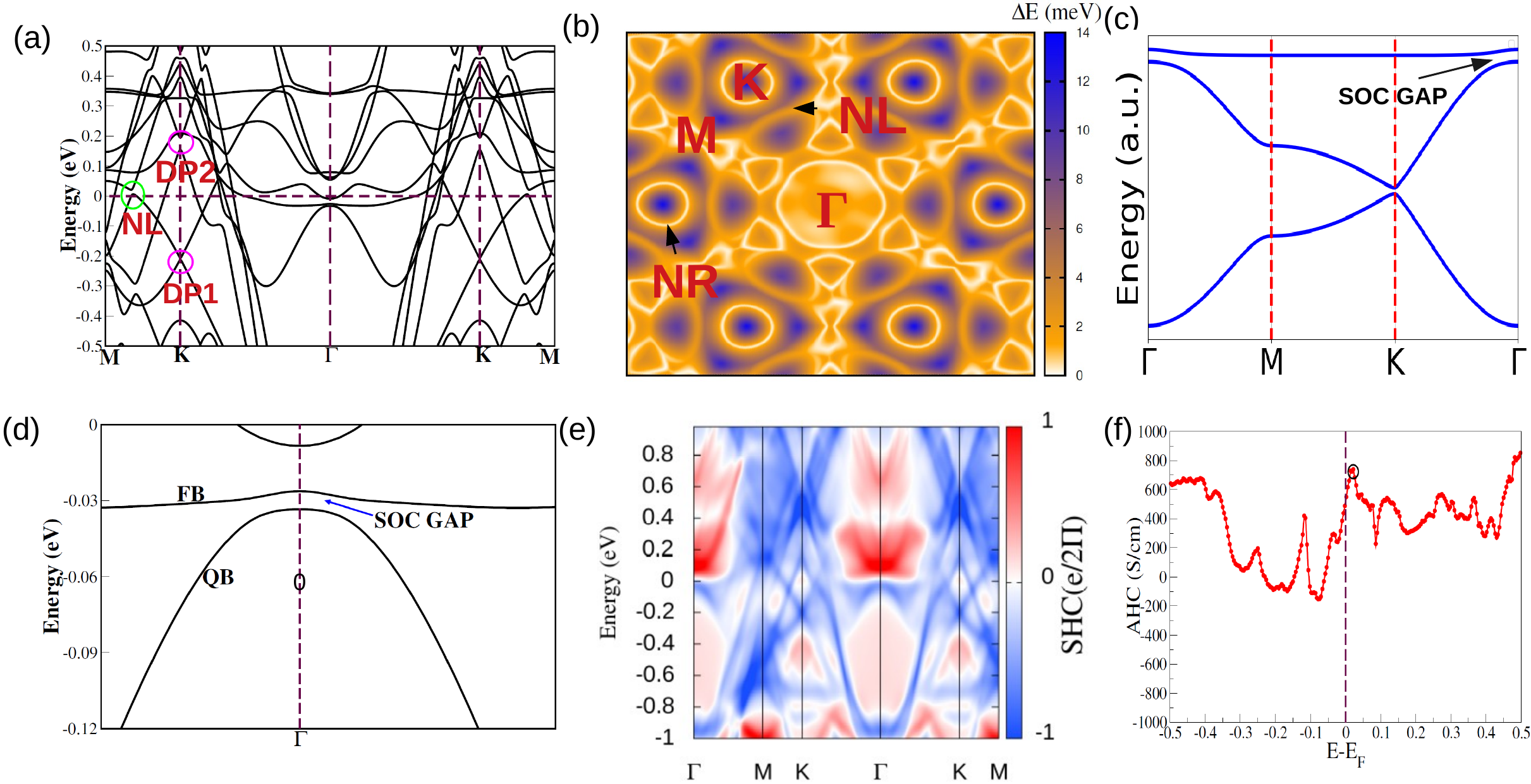}
    \caption{(a) The band structure of MgMn$_6$Sn$_6$
with SOC. (b) The energy gap between two crossing bands at \textit{E}$_F$ is shown, with the NR around the high-symmetry point \textit{K} marked by a white line. The gapped-out NL is indicated by a violet circle, and the gap Dirac point is represented by a blue section at the \textit{K} point. (c) Tight binding band structure in the presence of SOC. (f)The gap between the QB touching and the FB at the $\Gamma$ point in the presence of SOC. (d) The SHC of MgMn$_6$Sn$_6$ varies with high symmetry points. Integration of the in-plane momentum contributions up to a certain energy reveals a significant enhancement of SHC within the nontrivial SOC gap between the flat and Dirac bands (e) The energy(\textit{E}-\textit{E}$_F$ ) dependence of AHC for MgMn$_6$Sn$_6$, the gapped NL position is indicated by the black circle.}
    \label{fig4}
\end{figure*}

\subsection{Influence of SOC}

Topological properties are highly sensitive to SOC effects, in this section, we explore the impact of the SOC on the band dispersion resulting from the transition metal atom Mn.
Mn, with its significant SOC, is anticipated to have a substantial impact, particularly on the degenerate points, due to the mixing of up and down spin channels. From the tight-binding model, we have also noticed that with the coefficient of SOC even at 1/100th of the nearest neighbor hopping, the degenerate points are gapped out at \textit{K} and $\Gamma$. When we incorporate the second nearest neighbor hopping, the degeneracy breaks even at low SOC values, as illustrated in Fig.S6. 

On examining the electronic band structure in the presence of SOC, we find that the DP1 at the \textit{K} point exhibits a tiny bandgap opening (13 meV) in the presence of SOC, as shown Fig.\ref{fig4}(b). We have represented the energy gap where violet spots at the \textit{K} points represent the gap at DP1 as shown in Fig.\ref{fig4}(a)and this non-trivial Dirac fermion resides in the occupied state and is located closer to the \textit{E}$_F$, contributing significantly to the BC, depicted by a red spot in Fig.S8(a). Moreover, another crossing point, DP2 at the \textit{K} point, exhibits a wider band gap than DP1 opening (42 meV) in the presence of SOC. However, since it is distant from the \textit{E}$_F$, it will not impact the electron conduction. Nevertheless, by doping electrons into the system, we can access this larger gap.

In part A we have pointed out the degenerate point which forms the nodal line (NL). The NL loops in the \textit{k}$_z$ = 0 and \textit{k}$_z$ = $\pi$/c planes are protected by the M$_x$ mirror\cite{symmetry,sau2023,sau2024} symmetry and these NLs in the $\Gamma$-\textit{K}-\textit{H}-\textit{A} plane that links them. The six-fold rotational symmetry results in six NL and NL loops symmetrically distributed near the \textit{K} and \textit{L} points. However, due to the finite magnetization in the system, the time-reversal symmetry is broken, leading to the NLs beginning to gap out in the presence of SOC, although the gap size remains small ($<$ 14 meV) which is shown in Fig.\ref{fig4}(a). We have illustrated the energy gap of the NL in the \textit{k}$_z$=0 plane in Fig.\ref{fig4}(b), where a violet circle centered at \textit{k} denotes this gap. This gap induces a significant BC along this line, as depicted by a red speck in Fig.S8(a).

Another crossing point, indicated by an orange circle, remains ungapped even in the presence of SOC, as depicted in Fig.\ref{fig4}(a). The energy gap in the \textit{k}$_x$-\textit{k}$_y$ plane is illustrated in Fig.\ref{fig4}(b), where the white circular ring surrounding the K point forms the nodal ring (NR), forming Weyl points at certain k-points in momentum space. We have listed the Weyl points in Table~\ref{tab1} and also computed the normalized BC enclosing the coordinates of the points in the \textit{k}$_z$ = 0 plane. It is noted that the BC is produced by the Weyl point with chirality +1 as a source (outward flux in Fig.S7(a) and by the Weyl point with chirality -1 as a sink (inward flux in Fig.S7(b)). To verify the chirality of Weyl points, we conducted wannier charge centre (WCC) calculations. Fig.S7(c) illustrates the average WCC shifting from south to north when the Chern number of W$_1$ is positive, while Fig.S7(d) illustrates the average WCC shifting from north to south when the Chern number of W$_1$ is negative. We performed the same analysis for the other Weyl nodes and confirmed their chirality.
\begin{table}[hbtp]
\caption{\label{tab1}%
The Weyl points positions, Chern numbers, and the energy relative to the \textit{E}$_F$ of MgMn$_6$Sn$_6$.}
\begin{ruledtabular}
\begin{tabular}{ c c c c c c }
Weyl points &$k_x$ &$k_y$ & $k_z$ & Chern number& E-E$_F$\\
\hline
$W_1$&0.188&0.294 &0.0009  & -1 &-0.05\\
$W_1$&-0.239 &0.147 & 0.00002  & 1 & -0.08\\
$W_2$&-0.105&-0.060 &-0.0005 &  1 & -0.065\\
$W_2$&-0.188 &0.056 & 0.0003  &-1 & -0.083\\
$W_3$& -0.05 & -0.164 & 0.0009  & 1 & -0.1\\
$W_3$& -0.007 & 0.121 & 0.00009 & -1 & -0.12\\
\end{tabular}
\end{ruledtabular}
\end{table}
 In our investigation, we explored the impact of the SOC on the touching point of the quadratic band and FB, as illustrated in Fig.\ref{fig4}(c) from the tight-binding calculation in the presence of SOC with nearest neighbour hopping. We observed a gap opening at the contact point at $\Gamma$ as well as at the Dirac point which we have already discussed.  
 
 %Both the  quadratic band and the FB originates from the same orbital (dxz/dyz orbitals), a close examination of the energy distribution curves in Fig.\ref{fig3}(f) unmistakably demonstrates the emergence of a spin-orbit-induced gap at the point of quadratic touching.
The gap at the contact point of these two bands in MgMn$_6$Sn$_6$ measures 20 meV as shown in Fig.\ref{fig4}(d), which is smaller than the gap observed in CoSn (80 meV, \cite{flatgap}), yet it is closer to the \textit{E}$_F$. The SOC induced gap strongly suggests that the reported flat band at \textit{k}$_z$ = 0 has a nontrivial origin, endowing nonzero Z$_2$ invariant under the time-reversal breaking condition \cite{z2,z}. To derive the Z$_2$ index, the parities of the occupied bands at the time-reversal-invariant (TRI) moments are analyzed and six sets of Z$_2$ numbers were computed for them. If the system is in a Z$_2$-trivial state, this index will be even, and if not, odd. The Z$_2$ calculations of the six TRI planes can be utilized to obtain the Z$_2$ topological indices ($\nu_0$, $\nu_1\nu_2\nu_3$) using the formula:$\nu_0= (Z_2(k_{i}=0) + Z_2(k_{i}=0.5))\mod2 $ and $\nu_i= (Z_2(k_{i}=0.5))$. Consequently, MgMn$_6$Sn$_6$ is identified as a topological metal with a bulk Z$_2$ topological number of (1,011). This leads to the emergence of possible topologically nontrivial surface states. Surface bands can be observed between any pair of diametrically opposing points on the NR, as confirmed by the drumhead-like surface states in surface computations, which generally validate the NR state. The topological charges of the WPs are identified along the $\Gamma$-\textit{K}-\textit{M} direction. As depicted in Fig.S8(c), these WPs are further projected onto various surface momentum routes, resulting in the surface energy spectra exhibiting Fermi arcs for the top surface.

\subsection{Transport Properties}
The electronic structure of topological materials is intimately linked to their transport properties. In this section, we examine the transport properties of MgMn$_6$Sn$_6$, specifically exploring phenomena such as the spin Hall effect and anomalous Hall effect. The immediate detection of the SOC separation amid the Dirac and the flat band emphatically indicates the complex structure of the noted flat band at \textit{k}$_z$=0 plane.

To support the nontriviality of FB, we employ the DFT-based Wannier tight-binding model to examine the parity eigenvalue of the flat bands at the \textit{k}$_z$=0 plane using the Fu-Kane formula\cite{fu}. This examination results in a topological index Z$_2$=1 for the flat band, affirming their topological character. To illustrate how the nontrivial topology of the flat bands influences bulk properties, we additionally computed the band-resolved SHC for our compound. The constructed Wannier tight-binding model for MgMn$_6$Sn$_6$ enables us to conduct \textit{ab initio} calculations for the SHC and the k-resolved contributions from each band, utilizing the Kubo formula\cite{kubospin,kubospin1}. The in-plane momentum-resolved SHC primarily concentrates near the point where the degeneracy between the Dirac and flat bands is lifted, a feature connected to its topological nature. This connection is crucial for the formation of quantized SHC within the SOC induced gap, as depicted in Fig.\ref{fig4}(e).

The AHC is directly linked to the BC, resulting in a transverse anomalous velocity acquired by the electronic motion. Evaluating the intrinsic AHC involves applying linear response theory within the Kubo formalism\cite{Gradhand_2012}, focusing specifically on the AHC in the xy-plane by integrating the BC of the bands that are occupied in the whole BZ [Eq.\ref{eq1} and Eq.\ref{eq2}]. A 501 $\times$ 501 $\times$ 501 k grid is used to calculate intrinsic AHC using maximally localised Wannier orbitals. The spin polarised bands of the Mn kagome lattice, produce large intrinsic BC, due to the presence of the DP1, and gapped nodal line close to the \textit{E}$_F$. The variation of AHC with \textit{E}$_F$ is shown in Fig.\ref{fig4}(f). We found a substantial intrinsic AHC of roughly 500 S/cm at the \textit{E}$_F$.

\section{Summary}

In conclusion, we have calculated the electronic structure of the intriguing topological itinerant metal MgMn$_6$Sn$_6$, characterized by three-dimensional bilayer kagome Mn layers sandwiching a honeycomb Sn layer. We present comprehensive electronic properties using ab initio DFT and DMFT calculations. Utilizing the two-orbital Hubbard model Hamiltonian informed by insights from DFT calculations, we solve the model for a unit cell to analyze the many-body configuration of the magnetic ground state using ED. The electrons in the d-orbitals of the Mn atoms do not exhibit full spin polarization; instead, they form a frustrated many-body ground state with significant quantum fluctuations, resulting in a ground state that is a linear combination of many electronic configurations. Additionally, the hopping terms of electrons between Mn-Mn and Mn-Sn-Mn give rise to effective antiferromagnetic and ferromagnetic spin exchanges, respectively, inducing spin frustration in the system.

Our ab initio calculations suggest that in MgMn$_6$Sn$_6$, the nodal line in the \textit{k}$_z$=0 plane is protected by mirror symmetry. In the presence of SOC, gaps were found along the nodal line due to magnetic ordering, resulting in a large intrinsic BC. We have also demonstrated the presence of Dirac fermions at the Brillouin zone corner \textit{K}, VHS at the zone edge \textit{M}, and flat bands across the Brillouin zone of MgMn$_6$Sn$_6$ in the absence of SOC. The Dirac nodes gain a finite mass gap in the presence of SOC, contributing to the BC. Finite SOC breaks the degeneracy at the touching point at the zone center $\Gamma$ of the quadratic band and the flat band. To the best of our knowledge, the position of the flat band and the gap is the closest to the \textit{E}$_F$ ever reported in the literature.

The gap induced by SOC strongly suggests that the reported flat band possesses a nontrivial structure, resulting in a nonzero Z$_2$ invariant. Examining how the nontrivial topology of the flat bands influences bulk properties, we additionally computed the band-resolved spin Hall conductivity. The anomalous Hall conductivity is directly associated with the BC, leading to a transverse anomalous velocity during electronic motion. We observed a significant intrinsic AHC of approximately 500 S/cm at the \textit{E}$_F$. Altogether, the electronic properties of MgMn$_6$Sn$_6$ exhibit many interesting topological and frustrated magnetic properties. We demonstrate that the complexity of the magnetic and non-trivial topological properties of this system necessitates extensive many-body effects to accurately capture the nature of the many-body magnetic ground state. The robustness of the quantum phases requires sophisticated numerical methods such as DFT, ED, and DMFT computational techniques. We hope our study will influence further experimental analysis of this material, which may have potential applications in future devices, such as spin current generation, spin-orbit torques, spin Hall magnetoresistance and quantum computing (spintronic logic and memory devices).

\section*{Methods}

{\bf DFT Calculations:}
Our computational approach utilized the Vienna ab initio simulation package code, employing density functional theory (DFT). The generalized gradient approximation (GGA) was employed to approximate the exchange-correlation functional. We conducted first-principles calculations incorporating an effective Coulomb-exchange interaction U$_{eff}$ (U-J), where U and J represent the Coulomb and exchange parameters, respectively. To address the high electronic correlation effect of the Mn 3d electrons, we included an onsite Coulomb interaction parametrized with a Hubbard U$_{eff}$ = 3.0 eV \cite{dft+U}.The cutoff energy for expanding the wave functions into the plane-wave basis remained constant at 550 eV throughout the project. We utilized the Monkhorst-Pack scheme to sample the Brillouin zone in k-space for computations. The equilibrium structure served as the basis for the k-mesh, which was set at 10$\times$10$\times$6.The intrinsic Hall conductivity ($\sigma^{int}_{xy}$) was computed by integrating the z-component of Berry curvature ($\Omega^z$) over all occupied states across the Brillouin zone, with spin-orbit coupling (SOC) taken into account.

\begin{equation}
\label{eq1}
 \sigma_{xy} = -{\frac{e^2}{\hbar} \int\frac{d^{3}\textit{k}}{(2\pi)^3}\sum_{n}\Omega^z_{n}(\textit{k})f_n(\textit{k})}  \end{equation}
 $\Omega^{z}_n$  is Berry curvature and it can be written as\cite{xiao2010berry}
 \begin{eqnarray}
 \label{eq2}
\Omega^z_{n} = -2i \sum_{m \neq n} \frac{{\langle \psi_{n\textit{k}}|v_x|\psi_{m\textit{k}}\rangle} {\langle \psi_{m\textit{k}}|v_y|\psi_{n\textit{k}}\rangle}} {[E_{m}(\textit{k}) - E_{n}(\textit{k})]^2}
\end{eqnarray}

where $f_{n}(\textit{k})$ is the Fermi-Dirac distribution function, n is an index of the occupied bands, $E_{n}(\textit{k})$ is the eigenvalue of the $n^{\text{th}}$ eigenstate $\psi_{n}(\textit{k})$,$v_i$ = $\frac{1}{\hbar}\frac{\partial H(\textit{k})}{\partial \textit{k}_i }$ is the velocity operator along the i (i = x, y, z) direction.

{\bf DFT+DMFT Calculations:} For our DFT+DMFT calculations we are using the full-potential augmented plane-wave basis as implemented in the \textsc{wien2k} code package.\cite{wien2k}
For the \textsc{wien2k} calculations, we used the largest possible muffin-tin radii, and the basis set plane-wave cutoff was defined by ${R_{\text{min}}\!\cdot\!K_{\text{max}}} = 10$, where $R_{\text{min}}$ is the
muffin-tin radius of the O atoms. The consistency between the VASP and \textsc{wien2k} results have always been cross-checked.
DMFT calculations were performed using the TRIQS/DFTTools package \cite{aichhorn1, aichhorn2, aichhorn3} based on the TRIQS libraries\cite{triqs}.
%We perform DMFT calculations in a basis set of maximally localised Wannier functions (MLWF) Wannierized using Wannier90 \cite{Pizzi2020} and the wien2wannier \cite{wien2wannier} interface. 
Projective Wannier functions as implemented in the \textsc{dmftproj} module of TRIQS were employed to crosscheck the results and also to calculate the initial occupancy of the correlated orbitals. 
All five Mn $d$ orbitals have been taken into account in the correlated subspace. 
A projection window of $-12$\,eV to $+20$\,eV was chosen to take into account any hybridisation and charge transfer effects. 
The Anderson impurity model constructed by mapping the many-body lattice problem to a local problem of an impurity interacting with a bath was solved using the continuous-time quantum Monte Carlo algorithm in the hybridization expansion (CT-HYB)\cite{werner06} as implemented in the TRIQS/CTHYB package\cite{pseth-cpc}. For each DMFT step 1280000 cycles of warmup steps and 128000000 cycles of measures were performed for the quantum Monte Carlo calculations.
We performed one-shot DFT+DMFT calculations, using a fully localised limit (FLL) type double-counting correction\cite{helddc}. 
%Held type double-counting correction~\cite{anisimov93}
We use a fully rotationally-invariant Kanamori Hamiltonian parametrised by Hubbard $U$ and Hund's coupling $J_H$, where we set the intraorbital interaction to
$U'=U-2J_H$. 
For our DMFT calculations, we used $U$ value of 6\,eV and $J_H = 0.5$\,eV.  
The choice of U and J for DMFT calculations has been motivated by previous studies on Manganites excellent agreement between DMFT and experimental ARPES band structure, as well as band gaps, and $T_N$ have been demonstrated for a similar range of $U$ and $J_H$ values as well as prediction of experimental properties driven by correlations. \cite{held, lmo-cat, lmo-het}
Real-frequency spectra and real-frequency self-energy for band structure have been obtained using the maximum-entropy method of analytic continuation as implemented in the TRIQS/MAXENT application.\cite{maxent}

\section*{Data availability}

The authors declare that the data supporting the findings of this study are available within the paper and its 
supplementary information files.

\section*{Code availability}
The codes implementing the calculations of this study are available from the authors upon request.

\section*{acknowledgments}
M.K.  thanks SERB for financial support through Grant Sanction No.CRG/2020/000754. J.S. thanks  U.G.C  for financial support. S.S thanks DST-INSPIRE for financial support. H.B. thanks University of Dundee and University of Cambridge for support. DMFT calculations were performed on Sulis HPC cluster (EP/T022108/1) 

\section*{Contributions}
JS,MK and NK conceptualized the study. Numerical computations were done by JS, HB, and SS. All authors discussed the data and outcomes. All authors contributed to the manuscript's writing, which included their input.

\section*{Competing interests}
The authors declare no competing interests.
\bibliography{octupolar}

\end{document}

\title{SUPPLEMENTARY MATERIAL \\ Exploring magnetic and topological complexity in MgMn$_6$Sn$_6$: from frustrated ground states to nontrivial Hall conductivity} 
%\title{Octupolar magnetism in
%osmate double perovskites: Quadrupolar transverse fields and parasitic dipoles
%from impurity-induced strain}
%\title{Quadrupolar transverse fields and parasitic dipoles
%from impurity-induced strain in an octupolar magnet: Application to
%osmate double perovskites}
%\title{Suppression of Ising octupolar order from impurity-induced transverse fields: Application to osmate double perovskites}
%\title{Quadrupolar transverse fields and parasitic dipole moments
%on Ising octupolar order from impurity-induced strain: 
%Application to osmate double perovskites}
%$T_c$ suppression and parasitic dipole moments in the osmate double perovskites}
%\title{Inhomogeneous transverse fields on Ising octupolar order:\\ impurity induced strain in osmate double perovskites}
%Randomly localized transverse fields from local strains
\author{Jyotirmoy Sau}
\affiliation{Department of Condensed Matter Physics and Material Science,
S. N. Bose National Centre for Basic Sciences, JD Block, Sector III, Salt Lake, Kolkata 700106, India}

\author{Hrishit Banerjee}
\email{hbanerjee001@dundee.ac.uk}
\affiliation{School of Science and Engineering, University of Dundee, Scotland, UK}
\affiliation{Yusuf Hamied Department of Chemistry, University of Cambridge, Cambridge, UK}

\author{Sourabh Saha}
\affiliation{Department of Condensed Matter Physics and Material Science,
S. N. Bose National Centre for Basic Sciences, JD Block, Sector III, Salt Lake, Kolkata 700106, India}

\author{Nitesh Kumar }
\email{nitesh.kumar@bose.res.in}
\affiliation{Department of Condensed Matter Physics and Material Science,
S. N. Bose National Centre for Basic Sciences, JD Block, Sector III, Salt Lake, Kolkata 700106, India}
\author{Manoranjan Kumar}
\email{manoranjan.kumar@bose.res.in}
\affiliation{Department of Condensed Matter Physics and Material Science,
S. N. Bose National Centre for Basic Sciences, JD Block, Sector III, Salt Lake, Kolkata 700106, India}
%\date{\today}

% \pacs{75.25.aj, 75.40.Gb, 75.70.Tj}

% \tableofcontents 

% \vfill\eject

\begin{abstract}
	We here provide supplemental explanations and data on the following topics in relation to the main text: (I)Projected bands and total density of state (II) Constant energy contour at different DP (III) Hopping matrix (IV) DMFT self-energies (V) Tight binding model   (VI) Normalized BC of Weyl points with proper chirality (VII)Surface band dispersion.
\end{abstract}
\date{\today}
\maketitle
%\tableofcontents
%\appendix

\section{Supplementary Note 1: Projected bands with total density Of state}
\renewcommand{\thefigure}{S1}
\begin{figure}[h]
	\centering
	\includegraphics[width=0.95\textwidth]{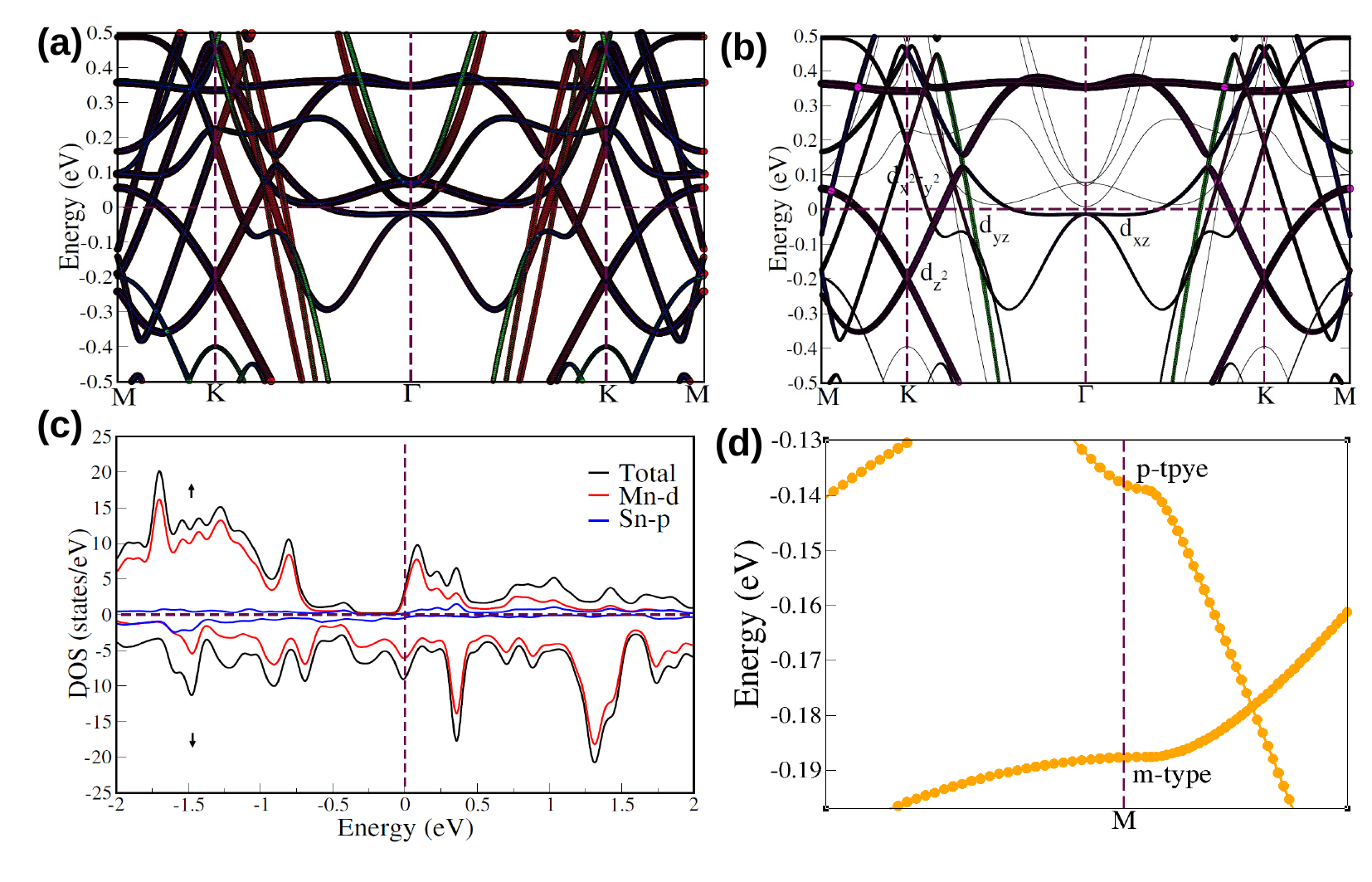}
\caption{(a)Mn, Sn, and Mg's projected orbital contributions to the band dispersion in MgMn$_6$Sn$_6$.   (b) Projected d-orbital contribution of Mn in the band dispersion in MgMn$_6$Sn$_6$. (c)The total DOS of MgMn$_6$Sn$_6$ is represented by the black curve, and the dos of Mn-d and Sn-p are represented by the red and blue curves, respectively.}
\label{fig:S2}
\end{figure}

\newpage

\section{Supplementary Note 2: Fermi surface}
\renewcommand{\thefigure}{S2}
\begin{figure}[h]
	\centering
	\includegraphics[width=0.5\textwidth]{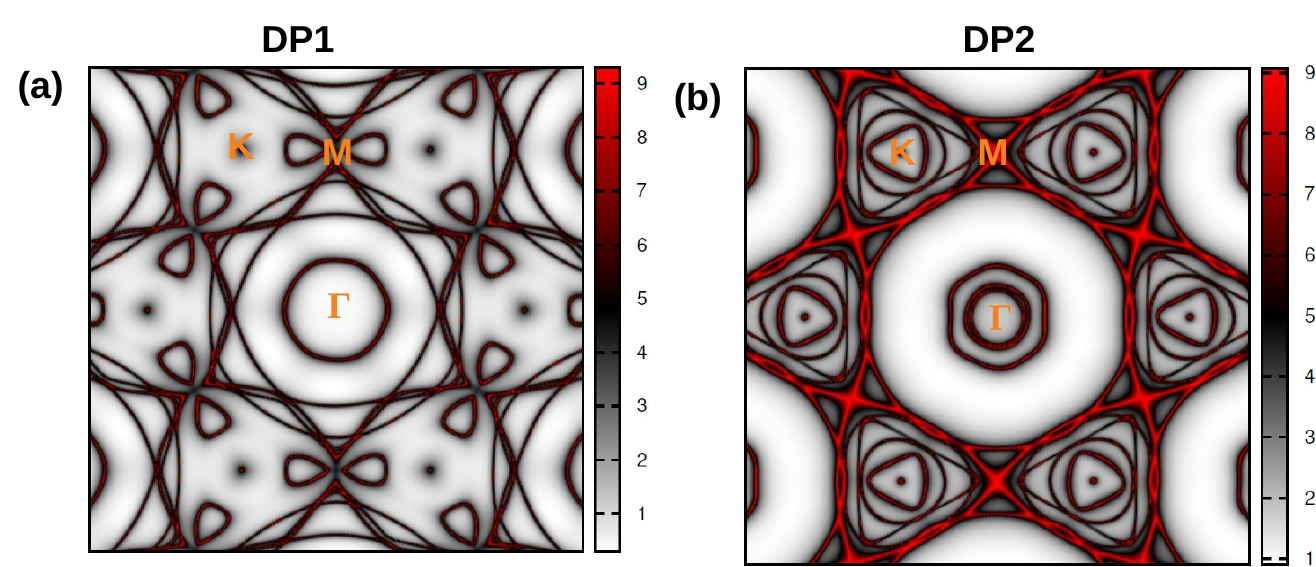}
 \caption{(a) Constant energy contour (CEC) at \textit{k}$_z$ = 0 plane estimated by DFT at DP1 (b) Constant energy contour (CEC) at \textit{k}$_z$ = 0 plane estimated by DFT at DP2 }
\label{fig:S1}
\end{figure}

\section{Supplementary Note 3: Hopping Matrix AND CONFIGARATIONS OF LOW SPIN STATE}
Here we have given the hopping parameters based of Wannier calculation for our material. The hopping matrix for Mn to Mn, which we took in our calculation is given by:
$t^{Mn-Mn}$= 
$\begin{pmatrix}
-0.1 & -0.1   \\
-0.1 & -0.1  \\
\end{pmatrix}.$

Hopping strength for each of the two orbital of Mn to single orbital of Sn is taken as, $t^{Mn-Sn}$=-0.8.\\

In the phase diagram presented in Fig. 3(c) of the main text, it is observed that as we increase the $J_H/U$ ratio for a fixed U, the system transitions from a low spin state (spin-1/2) to a high spin state (spin-5/2). The spin configuration in the low spin state is shown in Fig.S3 of the supplementary section. The most dominant configuration shows an antiferromagnetic alignment of spins between two different orbitals of each Mn site in the upper layer (UL) due to geometrical frustration. In the lower layer (LL), two Mn sites exhibit antiferromagnetic spin alignment between the spins in different orbitals at the same site, while the third Mn site has one vacant orbital and one orbital filled with an up-spin electron. In this configuration, the Sn site is doubly occupied.

The second most dominant configuration maintains the same spin alignment as the first configuration in the UL. However, in the LL , one Mn site has antiferromagnetic spin alignment between two different orbitals, another Mn site has one up-spin electron and one vacant orbital, and the third Mn site has one orbital occupied by a down-spin electron and the other orbital filled by two electrons with opposite spins. Here, the Sn orbital is filled with an up-spin electron.

\renewcommand{\thefigure}{S3}
\begin{figure}[h]
	\centering
	\includegraphics[width=.50\textwidth]{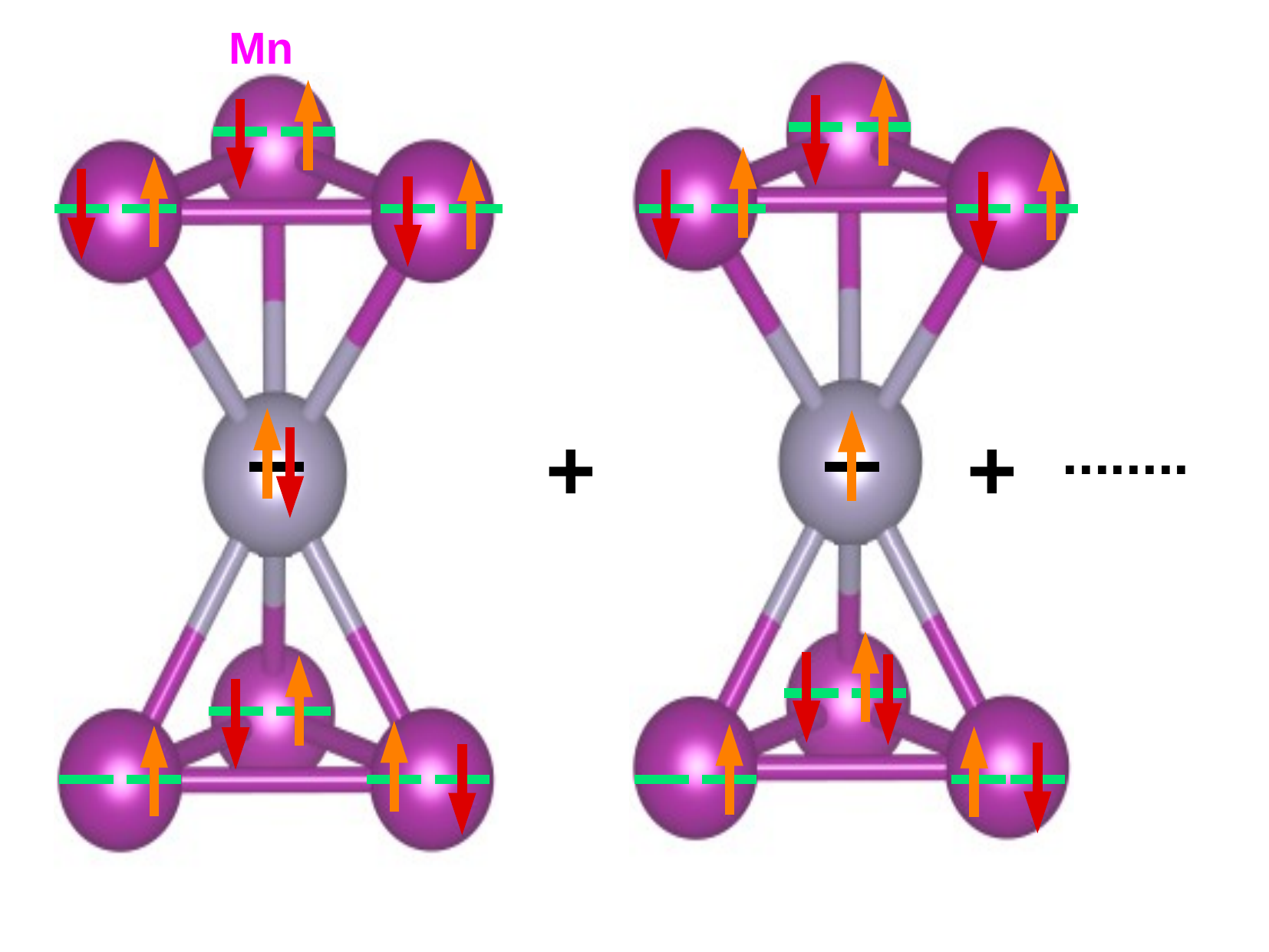}
 \caption{Two most dominant configurations in the low spin (spin 1/2) state.}
\label{fig:S1}
\end{figure}
In Fig.3(a) of the main text, we illustrate the two most dominant configurations in the ground state wave function for the high spin state.  The first configuration demonstrates that all orbitals of each atom are singly occupied due to the onsite Coulomb repulsion (U). In the LL of Mn atoms, spins tend to align in the same direction due to the large Hunds coupling ($J_H$). In contrast, in the UL, spins of two Mn atoms occupying different orbitals align in the same direction. However, these two Mn sites exhibit antiferromagnetic spin alignment to each other. The spins of another Mn atom in the UL, occupying different orbitals, align in the opposite direction. This antiferromagnetic exchange arises from the hopping of electrons between one Mn orbitals to the nearest Mn orbitals, and the triangular geometry leads to frustration. However, the Sn electron is in a down spin configuration due to the large hopping between Mn and Sn. This configuration has 12 replicas coming due to the symmetry in the system.
Another dominant configuration maintains ferromagnetic coupling in the LL triangle of Mn, similar to the first configuration, but for UL one electron from the d-orbital of the Mn-site is transferred to the Sn site making it double occupied. Consequently, Sn exhibits an effective zero magnetic moment. It has 36 degenerate replicas in the gs wave function.

Spin resolved ground state charge density for the degenerate ground state wave function in the high spin state is shown in Fig.S4 :
\renewcommand{\thefigure}{S4}
\begin{figure}
	\centering
	\includegraphics[width=1.00\textwidth]{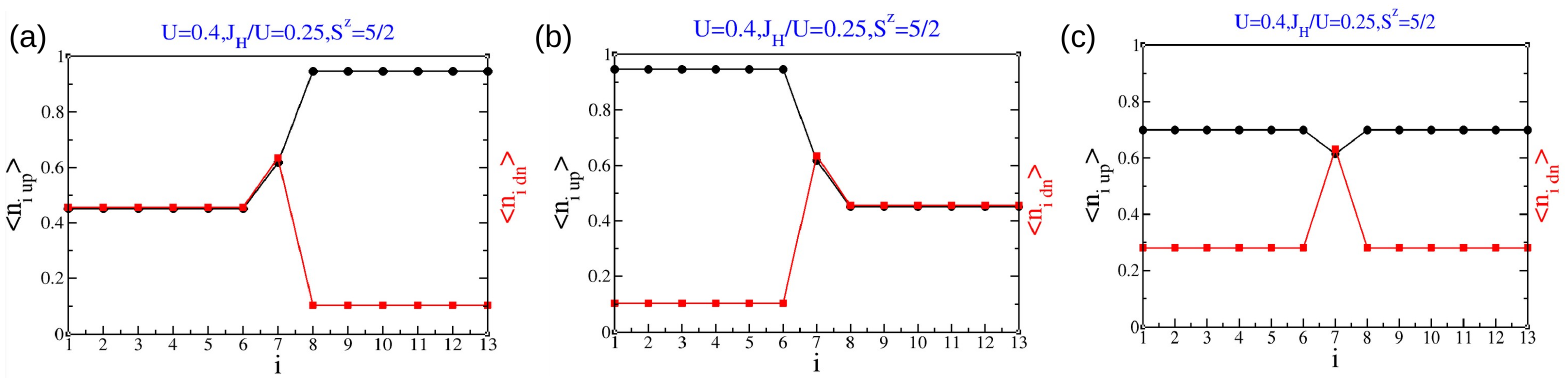}
 \caption{(a) and (b) correspond to the up and down spin charge densities shown by black and red lines for all the orbitals of each site for two degenerate ground state wave functions respectively. Fig (c) is the average up and down spin charge density for these two degenerate wave functions.}
\label{fig:S2}
\end{figure}
\newpage
\section{Supplementary Note 4: DMFT Self-Energies}
In this figure, we plot the Im $\Sigma$ (Self-energy) as a function of $\omega$ the Matsubara Frequencies. We observe a large finite value of Im $\Sigma$ as $\omega \rightarrow 0$, which confirms the incoherent scattering behaviour in this material. The variation in Self-energy shows the characteristic behaviour of non-Fermi liquid-type strongly correlated states. This incoherent scattering behaviour results in the lifetime broadening of the DMFT bands shown in Fig. 2(c) and (d) as shown in the main text. The split between up and down spin channel self-energies also shows the magnetic behaviour at $\beta=150$. A strong presence of orbital selective correlations is seen with large self-energy intercepts for some orbitals and spin channels compared to others. 
\renewcommand{\thefigure}{S5}
\begin{figure}[h]
\centering
\includegraphics[width=0.65\linewidth]{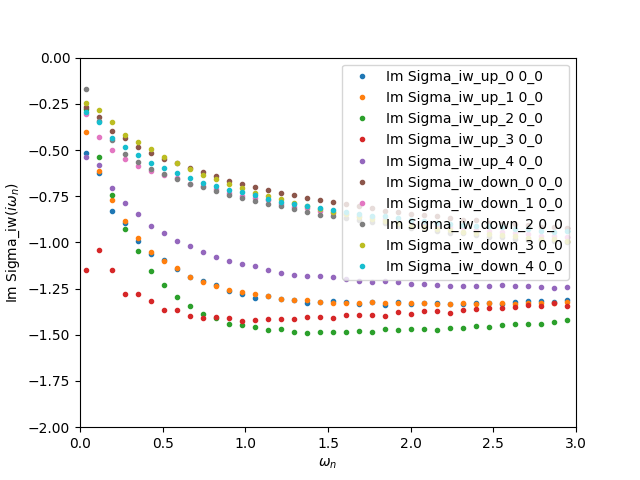}
\caption{ Figure showing Im $\Sigma$ as a function of $\omega$ plotted for $\beta=150eV^{-1}$, for the diagonal elements for 5 different Mn orbitals and two spin channels. } 
\label{fig8}
\end{figure}
\newpage
\section{Supplementary Note 5: Tight-binding model}

We consider nearest and next-nearest neighbor hopping on a kagome lattice with spin-orbit interactions, represented by the following Hamiltonian:
\begin{equation}
H = -t_1 \sum_{\langle i,j \rangle \sigma} c^{\dagger}_{i\sigma} c_{j\sigma} + i\lambda_1 \sum_{\langle i,j \rangle \alpha\beta} (\mathbf{E}_{ij} \times \mathbf{R}_{ij}) \cdot \boldsymbol{\sigma}_{\alpha\beta} c^{\dagger}_{i\alpha} c_{j\beta} 
- t_2 \sum_{\langle\langle i,j \rangle\rangle \sigma} c^{\dagger}_{i\sigma} c_{j\sigma} + i\lambda_2 \sum_{\langle\langle i,j \rangle\rangle \alpha\beta} (\mathbf{E}_{ij} \times \mathbf{R}_{ij}) \cdot \boldsymbol{\sigma}_{\alpha\beta} c^{\dagger}_{i\alpha} c_{j\beta}
\end{equation}

where $c^{\dagger}_{i\sigma}$ creates an electron with spin $\sigma$ on the site $\mathbf{r}_i$ on the kagome lattice. Here $\langle i,j \rangle$ denotes nearest neighbors and $\langle\langle i,j \rangle\rangle$ next-nearest neighbors. The second and fourth terms describe spin-orbit interactions which preserve time-reversal invariance. $\mathbf{R}_{ij}$ is the distance vector between sites $i$ and $j$ and $\mathbf{E}_{ij}$ the electric field from neighboring ions experienced along $\mathbf{R}_{ij}$.

We first study just nearest-neighbor hopping so t$_2$ =
$\lambda_2$ = 0. In momentum space, Eq.1 becomes

\begin{equation}
H(\mathbf{k}) = -2t_1 
\begin{pmatrix}
0 & \cos k_1 & \cos k_2 \\
\cos k_1 & 0 & \cos k_3 \\
\cos k_2 & \cos k_3 & 0
\end{pmatrix}
\pm i 2 \lambda_1
\begin{pmatrix}
0 & \cos k_1 & -\cos k_2 \\
-\cos k_1 & 0 & \cos k_3 \\
\cos k_2 & -\cos k_3 & 0
\end{pmatrix}
\end{equation}

where $\mathbf{a}_1 = \hat{x}$, $\mathbf{a}_2 = (\hat{x} + \sqrt{3}\hat{y})/2$, $\mathbf{a}_3 = \mathbf{a}_2 - \mathbf{a}_1$, and $k_n = \mathbf{k} \cdot \mathbf{a}_n$. We use units where the hopping parameter $t_1 = 1$ and $\lambda_1$ = 0.01. The $+$ ($-$) sign refers to spin up (down) electrons; from here we focus on just the spin up electrons.\\

For a more realistic scenario, we next include second-nearest neighbor hopping. This gives us additional terms in the Hamiltonian:

\begin{equation}
H(\mathbf{k}) = -2t_2 
\begin{pmatrix}
0 & \cos(k_2 + k_3) & \cos(k_3 - k_1) \\
\cos(k_2 + k_3) & 0 & \cos(k_1 + k_2) \\
\cos(k_3 - k_1) & \cos(k_1 + k_2) & 0
\end{pmatrix}
+ i 2 \lambda_2
\begin{pmatrix}
0 & -\cos(k_2 + k_3) & \cos(k_3 - k_1) \\
-\cos(k_2 + k_3) & 0 & -\cos(k_1 + k_2) \\
\cos(k_3 - k_1) & -\cos(k_1 + k_2) & 0
\end{pmatrix}
\end{equation}

We use units where the hopping parameter $t_2 = 0.3$ and $\lambda_2$ = 0.008.
\renewcommand{\thefigure}{S6}
\begin{figure}[h]
	\centering
	\includegraphics[width=0.45\textwidth]{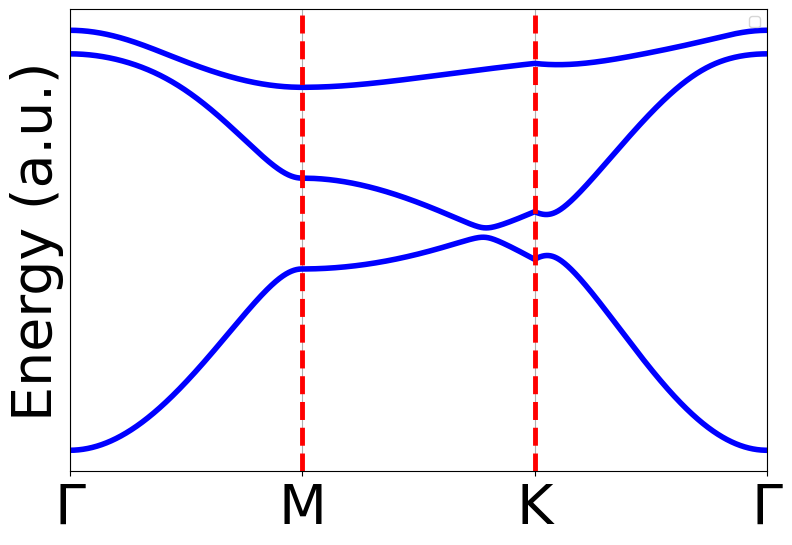}
 \caption{Dispersion of energy bands in the presence of spin-orbit coupling (SOC) with second-nearest neighbor hopping. }
\label{fig:S1}
\end{figure}

\section{Supplementary Note 6: Chirality of weyl points }
\renewcommand{\thefigure}{S7}
\begin{figure}[h]
	\centering
	\includegraphics[width=0.85\textwidth]{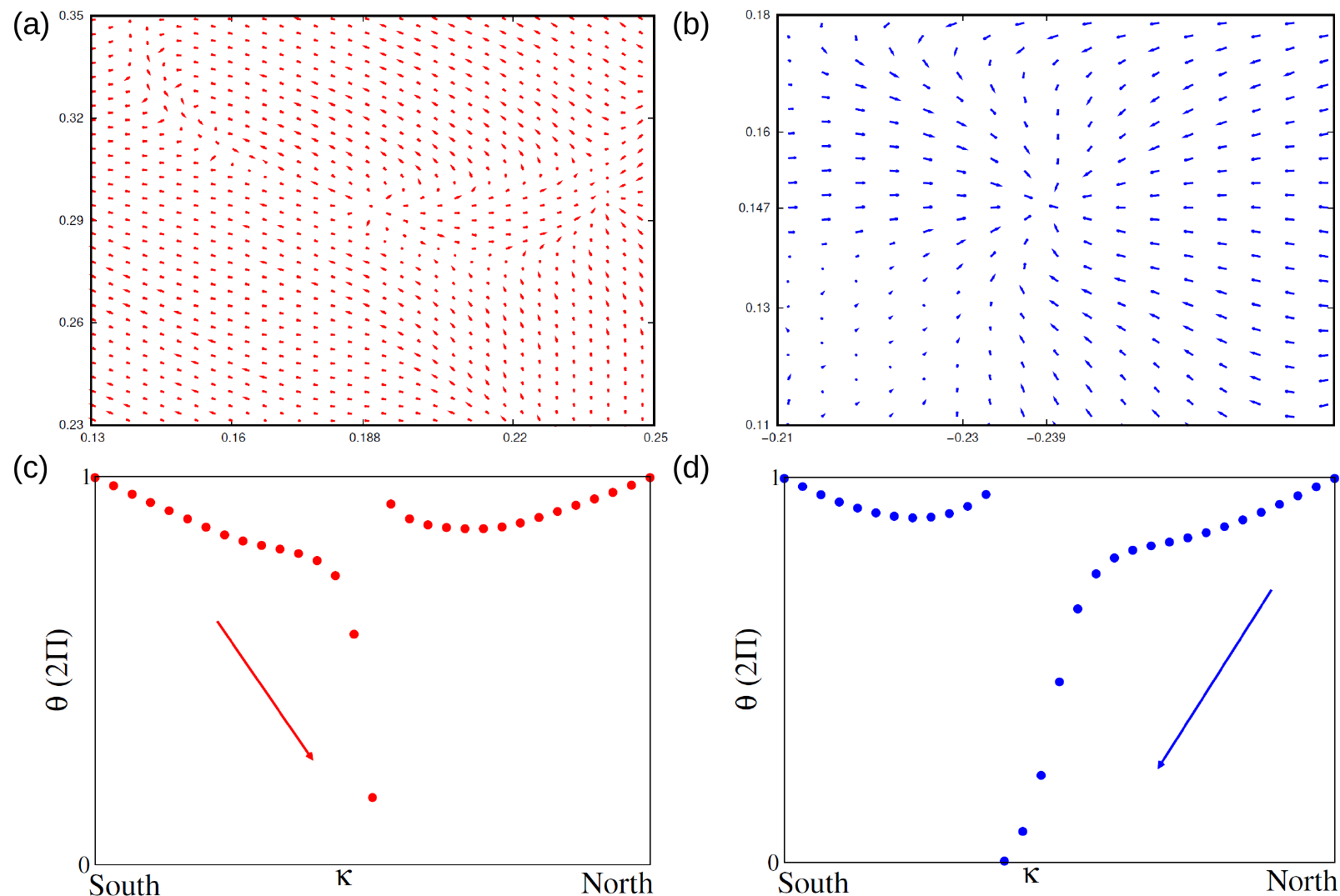}
\caption{ MgMn$_6$Sn$_6$ with SOC normalized Berry curvatures is shown for Weyl point W$_1$ (a) Source type (W$_1+$) which is indicated by the outward red arrow from a marked black circle. (b) Sink type (W$_1-$) which is indicated by an inward blue arrow marked black circle. (c)The average position of the Wannier charge centre corresponds to +1 chern number. (d) average position of the Wannier charge centre corresponds to a $-1$ chern number.}
\label{fig:S3}
\end{figure}

\newpage
\section{Supplementary Note 7: Berry Curvature with nontrivial surface state}
\renewcommand{\thefigure}{S8}
\begin{figure}[h]
	\centering
	\includegraphics[width=.950\textwidth]{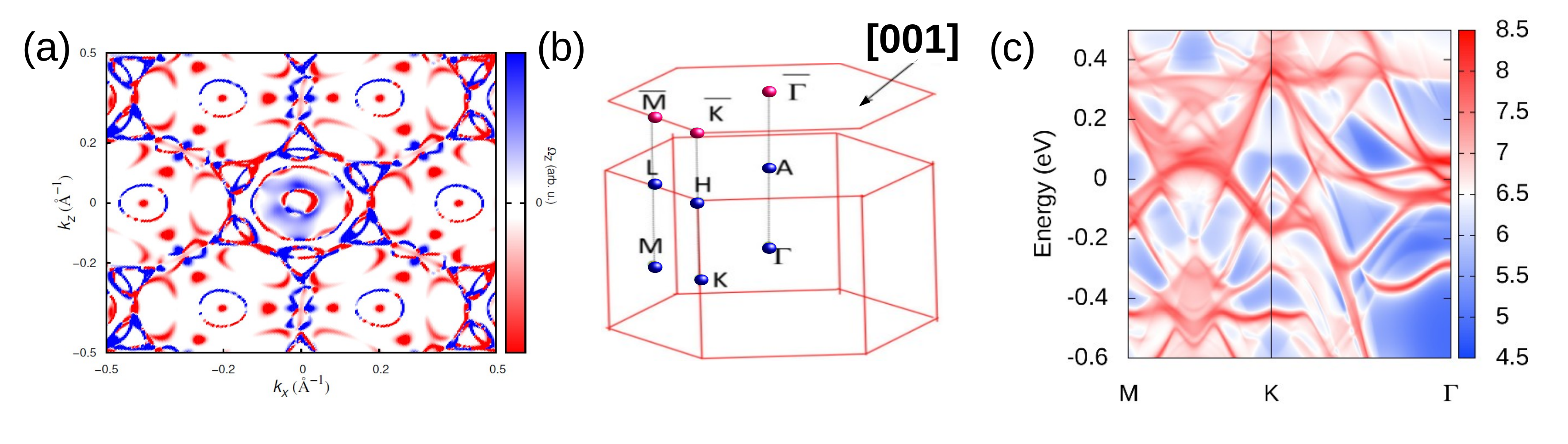}
\caption{(a) BC distribution in \textit{k}$_x$-\textit{k}$_y$ plane. (b) Brillouin Zone with high symmetry point (c) displays the surface band dispersion for the slab (001).}
\label{fig:S4}
\end{figure}

\bibliography{octupolar}